\RequirePackage{fix-cm}
\documentclass[twocolumn]{svjour3}          % twocolumn
\smartqed  % flush right qed marks, e.g. at end of proof
\usepackage{graphicx}
\graphicspath{{figures/}} 
\usepackage{amsmath}
 \journalname{Med Biol Eng Comput}
\begin{document}

\title{Vinamax: a macrospin simulation tool for magnetic nanoparticles
%\title{Sensitivity analysis of 1-dimensional magnetic nanoparticle mappings with electron paramagnetic resonance%\thanks{Grants or other notes
%\title{of Sensitivity analysis of 1-dimensional magnetic nanoparticle mapping with electron paramagnetic resonance of Determination of the limiting factors of 1-dimensional magnetic nanoparticle mapping with electron paramagnetic resonance%\thanks{Grants or other notes
}

\author{Jonathan Leliaert \and Arne Vansteenkiste \and Annelies Coene \and \\ Luc Dupr\'e \and Bartel Van Waeyenberge }


\institute{J. Leliaert \and A. Vansteenkiste \and B. Van Waeyenberge \at
             Department of Solid State Sciences, Ghent University, Krijgslaan 281/S1, 9000 Ghent, Belgium\\
              \email{jonathan.leliaert@ugent.be}           %  \\
           \and
	   J. Leliaert \and A. Coene \and L. Dupr\'e \at
              Department of Electrical Energy, Systems and Automation, Ghent University, Sint-Pietersnieuwstraat 41, 9000 Ghent, Belgium \\
}

\date{Received: date / Accepted: date}

\maketitle
%%TO DO%%%%%%%%%%%%%%%%%%%%%%%%%%%%%%%%%%%%%%%%%%%%%%%%%%%%%%%%%%%%%%%%%%%%%%%%%%%%%%%%%%%%%%%%%%%%%%%%%%%%%%%%%%%%%%%%%%%
%\textbf{algemeen: These regular publications describe original contributions to the advancement of Medical and Biological Engineering and Computing. Recommended length: not exceeding 6000 words including title page, abstract, text, references, tables, and figure legends. The main text of Original Research Articles should contain the following separate sections: Introduction, Methods, Results, Discussion. The maximum number of figures and tables together is 8. 9 blz?}
%%%%%%%%%%%%%%%%%%%%%%%%%%%%%%%%%%%%%%%%%%%%%%%%%%%%%%%%%%%%%%%%%%%%%%%%%%%%%%%%%%%%%%%%%%%%%%%%%%%%%%%%%%%%%%%%%%%%%%%%%%

\begin{abstract}
We present Vinamax, a simulation tool for nanoparticles that aims at simulating magnetization dynamics on very large timescales. To this end, each individual nanoparticle is approximated by a macrospin. Vinamax numerically solves the Landau-Lifshitz equation by adopting a dipole approximation method, while temperature effects can be taken into account with two stochastic methods. It describes the influence of demagnetizing and anisotropy fields on magnetic nanoparticles at finite temperatures in a space and time-dependent externally applied field. Vinamax can be used in biomedical research where nanoparticle imaging techniques are under developement. 
%Insert your abstract here. Include keywords, PACS and mathematical
%subject classification numbers as needed.
%\PACS{75.10.Hk \and 75.75.Jn \and 75.78.Cd}
\end{abstract}

\section{Introduction}
In recent years, many biomedical applications based on nanotechnology \cite{LIM-10} in general and magnetic nanoparticles\cite{PAN-03,PAN-09} specifically have emerged. Examples of applications under development are disease detection \cite{KIR-03}, targeted drug delivery \cite{ALE-00} and hyperthermia \cite{JOH-07}. All of these depend upon an accurate knowledge of the spatial distribution of the magnetic particles \cite{COE-12,EIC-12} and thus require a nanoparticle imaging technique \cite{BAU-08,LLA-10}. Some promising imaging techniques are Magnetic Particle Imaging \cite{GLE-05}, Magnetorelaxometry \cite{WIE-12} and Electron Paramagnetic Resonance \cite{COE-13}. While each method has its distinct advantages, none of them is able to quantitatively reconstruct the spatial particle distribution \textit{in vivo}.
Inadequate knowledge of the spatial distribution of the nanoparticles results in suboptimal applications and even a decrease in patient safety and comfort \cite{COE-13}. The development of an adequate technique is challenging, in part, due to an insufficient understanding of the collective magnetic behaviour of the particles. %As numerical simulations form an important aspect of modern research, we argue that 
However, numerically investigating these particles from first principles (i.e. micromagnetically \cite{BRO-63a}) can lead to an understanding of this collective behaviour and consequently an improved performance of aforementioned applications. To the best of our knowledge there exists no simulation software that both is accurate on the smallest timescales at which micromagnetic dynamics take place, and still is able to simulate the large timescales involved in experiments. Therefore, we have developed Vinamax\cite{VIN-14}: a broad simulation tool in which individual nanoparticles are represented by single macrospins. Vinamax numerically solves the Landau-Lifshitz equation \cite{LAN-35} and considers demagnetizing and anisotropy fields. It also takes into account externally applied fields that can be space and time-dependent. To be able to simulate large ensembles of nanoparticles the demagnetizing interaction is solved efficiently using a dipole approximation method \cite{TAN-00}. This contrasts approaches in which the demagnetizing interaction is not taken into account \cite{BAB-10} or cut off after a short distance, e.g. when considering aggregates of nanoparticles \cite{AND-12}. Additionally, thermal effects \cite{BRO-63,EVA-12} can be taken into account by two different approaches: first, a stochastic field term can be added to the effective field or second, magnetic nanoparticles are switched at stochastic time intervals, equivalent to the first approach.\\

The paper is organized as follows. In Section \ref{methods} the Landau-Lifshitz equation with the addiational stochastic term is described in detail together with the algorithm used to calculate the demagnetizing interaction. Section \ref{validation} demonstrates the validity of the software by comparing simulation results to the well-established micromagnetic software \textsc{MuMax3}\cite{VAN-11}. In Section \ref{application} the equivalency of both approaches to include thermal effects is demonstrated and a challenging simulation is considered in which one million particles are simulated for one second, as is necessary to simulate a magnetorelaxometry experiment. Finally, some concluding remarks about Vinamax are given in Section \ref{conclusions}. 

\section{Methods}
\label{methods}
\subsection{Micromagnetic theory}
In the micromagnetic framework the magnetization is described as a continuum vector field $\textbf{M}(\textbf{r},t)$. In the following the space- and time dependence of the magnetization vector field is no longer explicitly shown in the equations. The nanoparticles under consideration are assumed to be uniformly magnetized\cite{NOW-05} (i.e. their size is sufficiently small to be single domain particles). Because all spins within each particle lie parallel to each other, Vinamax can further simplify the continuum approximation by describing every nanoparticle as one single macrospin. This implies that the exchange interactions do not have to be evaluated. The magnetic dynamics of each nanoparticle are described by the Landau-Lifshitz equation (\ref{ll-eq}):
\begin{equation}
	\tau=\frac{d\mathbf{m}}{dt}=-\frac{\gamma_0}{1+\alpha^2}\left(\mathbf{m}\times\mathbf{B}_{\text{eff}}+\alpha\mathbf{m}\times\mathbf{m}\times\mathbf{B}_{\text{eff}}\right)
	\label{ll-eq}
\end{equation}
In this equation, the torque $\tau$, due to the effective field $\mathbf{B}_{\text{eff}}$ felt by each nanoparticle, equals the time derivative of $\mathbf{m}$, where $\mathbf{m}$ denotes the magnetization vector normalized with respect to the saturation magnetization ($\textbf{M}=\textbf{m}$M$_{\text{sat}}$). Furthermore, $\gamma_0$ denotes the gyromagnetic ratio 1.7595 $\times 10^{11}$ rad/Ts and $\alpha$ is the dimensionless Gilbert damping constant. Eq. (\ref{ll-eq}) is solved numerically by timestepping it with timestep $\Delta$t, which is described in further detail in section \ref{timestep}.\\

The field $\mathbf{B}_{\text{eff}}$ is the sum of the different effective field terms that influence the magnetization:
\begin{equation}
\mathbf{B}_{\text{eff}}=\mathbf{B}_{\text{ext}}+\mathbf{B}_{\text{anis}}+\mathbf{B}_{\text{demag}}+\mathbf{B}_{\text{therm}}
\label{eq-beff}
\end{equation}
The different terms in equation (\ref{eq-beff}) are described in more detail below.
\subsubsection{External field}
$\mathbf{B}_{\text{ext}}$ is an externally applied field that can be both space and time-dependent.
\subsubsection{Anisotropy field}
In Vinamax the nanoparticles are assumed to have uniaxial anisotropy which is the case for the iron-oxide nanoparticles used in biomedical applications\cite{PAN-03}. The field affecting a particle due to this anisotropy is given by equation (\ref{eq-anis}): 
\begin{equation}
\mathbf{B}_{\text{anis}}=\frac{2K_{u1}}{M_{\text{sat}}}\left(\mathbf{m}\cdot\mathbf{u}\right)\mathbf{u}
\label{eq-anis}
\end{equation}

The anisotropy constant $K_{u1}$ and the anisotropy axis $\mathbf{u}$ can be chosen freely. $\mathbf{u}$ can be set to a predefined direction or to uniformly distributed random directions for all nanoparticles. To this end, two random numbers $\phi$ and $\theta$ are generated with the following distribution:
\begin{eqnarray}
	&\phi = 2\pi\rho_1\\
	&\theta = 2\arcsin(\sqrt\rho_2)
\end{eqnarray}
where $\rho_{1,2}$ denote two uncorrelated, uniformly distributed random numbers in the interval $\left[0.0,1.0\right)$.
These spherical coordinates are then mapped to their cartesian counterparts using the well-known relations below:
\begin{eqnarray}
	&x=\sin(\theta)\cos(\phi)\\
	&y=\sin(\theta)\sin(\phi)\\
	&z=\cos(\theta)
\end{eqnarray}

\subsubsection{Demagnetizing field}
The demagnetizing field (or the magnetostatic interaction) results from the dipole-dipole interaction of the different particles. Its contribution to the effective field is given by equation (\ref{eq-demag}), where $i$ loops over all particles, $\mu_0=4\pi\times 10^{-7}$ T/Am, $\mathbf{r}_i$ denotes the distance from each particle to the point at which the demagnetizing field is evaluated, and $V_i$ is the volume of each particle.   
\begin{equation}
	\mathbf{B}_{\text{demag}}=\mu_0\sum_{i}V_i{M_{\text{sat},i}\left[3\frac{\left(\mathbf{m}_i\cdot\mathbf{r}_i\right)\mathbf{r}_i}{r_i^5}-\frac{\mathbf{m}_i}{r_i^3}\right]}
\label{eq-demag}
\end{equation}
\subsubsection{Dipole approximation method}
To speed up the evaluation of the demagnetizing field, we have implemented a dipole approximation method. This method is based on a multipole approximation described in Ref. \cite{TAN-00}. We will briefly describe our adapted implementation, but for a detailed explanation we refer to Ref. \cite{TAN-00}.\\

In Vinamax, the user has to define a \textit{world}, which is a cube that encloses all the particles in the simulation. This cube is then subdivided into 8 subnodes, which are further subdivided until every node contains at most one particle. Within this tree, we then calculate the \textit{center of magnetization} once for every node. The center of magnetization (CM) is the center of mass of the particles in the node weighted with the magnetic moment of each particle. This is shown in eq. (\ref{eq-com}), where $\sum_i$ denotes a sum over all particles in the node, and $\mathbf{r}_i$ is the position of particle $i$. 

\begin{equation}
	\mathbf{R}_{\text{CM}}=\frac{1}{\sum_{i}V_iM_{\text{sat},i}}\sum_{i} \mathbf{r}_iV_iM_{\text{sat},i}
	\label{eq-com}
\end{equation}
In the evaluation of the total demagnetizing field we take into account the contribution per node of the field $\mathbf{B}_{\text{d}}$ as if all the particles in a node were compressed into one dipole in the CM of that node, using eq. (\ref{eq-dip}), where $\mathbf{r}$ denotes the distance between the CM of the node and the point at which the field is evaluated. Before every timestep, the magnetization $\mathbf{M}=\sum_i V_iM_{\text{sat},i}\textbf{m}_i$ of each node is updated based on the magnetization of every particle within the nodes.\\

\begin{equation}
	\mathbf{B}_{\text{d}}=\mu_0\frac{3(\mathbf{M}\cdot\mathbf{r})\mathbf{r}-\mathbf{M}}{r^3}
	\label{eq-dip}
\end{equation}
 
\begin{figure}
%\centering
\includegraphics[width=8.6cm]{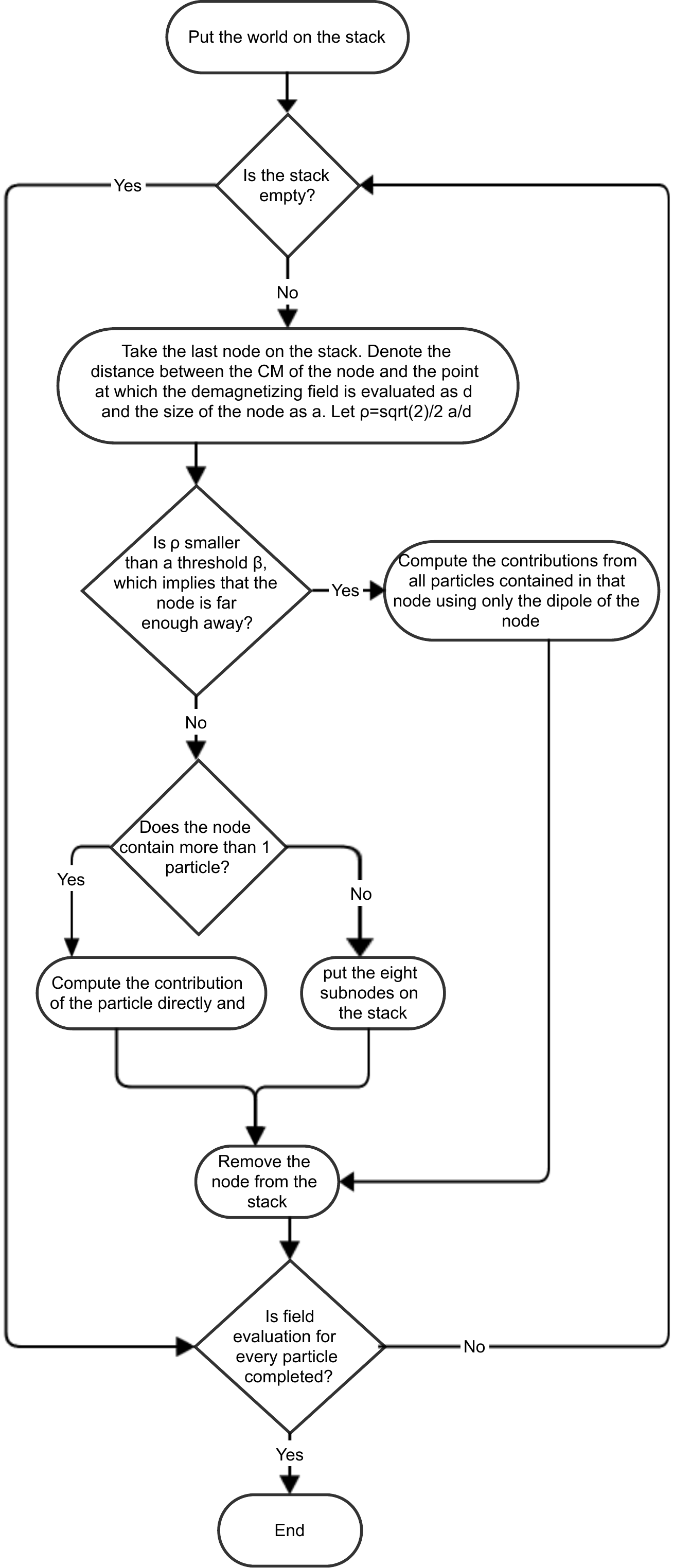}
\caption{\label{Fig_flowchart} The algorithm used to evaluate the demagnetizing field, adapted from Ref. \cite{TAN-00}.}
\end{figure}

For every particle, the demagnetizing field working on it is evaluated with the algorithm described in Fig. \ref{Fig_flowchart}, copied from Ref. \cite{TAN-00} with minor adaptations to our implementation.\\
%
%\begin{itemize}
%	\item Step 1. Put the \textit{world} on the stack;
%	\item Step 2. If the stack if empty, go to Step 6; otherwise
%	\item Step 3. Take the last node on the stack. Denote the distance between the CM of the node and the point at which the demagnetizing field is evaluated as $d$, and the size of the node as $a$. Let $\rho=\frac{\sqrt2}{2}\frac{a}{d}$. If $\rho$ is smaller than a threshold $\beta$, which implies that the node is far enough away, compute the contributions from all particles contained in that node using only the abovementioned dipole of the node and remove the node from the stack; otherwise
%	\item Step 4. If the node contains only one particle, compute the contribution of the particle directly and remove the node from the stack; otherwise
%	\item Step 5. Remove the node from the stack, put its eight subnodes on the stack, and go to Step 2;
%	\item Step 6. If the field evaluation for every particle is completed, go to Step 7; otherwise go to Step 1 and start the field evaluation for the next particle.
%	\item Step 7. End
%\end{itemize}

This algorithm only requires work of $\mathcal{O}(N log(N))$, while the direct pairwise evaluation of the demagnetizing field (from now on called \textit{brute force method}) between $N$ particles (eq. (\ref{eq-demag})) scales as $\mathcal{O}(N^2)$. These statements are validated in section \ref{validation}.

\subsubsection{Stochastic thermal field}
\label{brownfield}
The effective field $\mathbf{B}_{\text{eff}}$ includes a stochastic term $\mathbf{B}_{\text{therm}}$ to take thermal fluctuations \cite{COF-12} into account. Brown\cite{BRO-63} has worked out the detailed properties of this stochastic field for a single domain particle with the use of the fluctuation-dissipation theorem, which led to eq. (\ref{eq-therm}). % The statistical properties of each cartesian component of the stochastic vector field can be summarized as\cite{LOP-12}:
%\begin{eqnarray}
%	\label{eq-therm1}
%	&\langle\mathbf{B}_{\text{therm}}(t)\rangle=0\\
%	\label{eq-therm2}
%	&\langle\mathbf{B}_{i,thermal}(t)\mathbf{B}_{j,thermal}(t')\rangle=D\delta(t-t')\delta_{ij}\\
%	&D=\sqrt{\frac{2kT\alpha}{\gamma_0 M_{\text{sat}} Vdt}}
%\end{eqnarray}
%where $\langle\quad\rangle$ denotes a statistical average, $t$ and $t'$ are different points in time and $i$ and $j$ are indices denoting the different particles. Eq. (\ref{eq-therm1}) means that the thermal fluctuations have mean 0 while the Kronecker deltas in eq. (\ref{eq-therm2}) respectively mean that the thermal fluctuations are uncorrelated in time and between the different particles. Furthermore, the different components of the stochastic vector field are also uncorrelated.\\

Because in Vinamax each particle is approximated by one single macrospin, it is particularly suitable to use Brown's theory. So the equation we implemented reads 
\begin{equation}
	\mathbf{B}_{\text{therm}}(t)=\eta(t)\sqrt{\frac{2k_BT\alpha}{\gamma_0 M_{\text{sat}} V \Delta\text{t}}}
\label{eq-therm}
\end{equation}
where $\eta(t)$ denotes a random vector whose components are normal distributed random numbers with mean 0 that are uncorrelated in space and time, $\Delta$t is the timestep and $k_B$ is the Boltzmann constant.\\

\subsubsection{Stochastic switching}
\label{jumpnoise}
The approach with the stochastic field of section \ref{brownfield} has a large computational cost. Therefore, also a faster implementation based on stochastic switching of the nanoparticles is presented. This approach, however has the drawback that it is only valid in a constant field.\\

For a single domain particle with uniaxial anisotropy eq. (\ref{eq-therm}) leads to switching between two states with minimal energy, separated by an energy barrier of height $\Delta\text{E}=K_{\text{u1}}V$ with a switching rate $f$:
\begin{equation}
	f=f_0\exp{\left(\frac{-\Delta\text{E}}{k_BT}\right)}
\label{eq-f}
\end{equation}
In the high barrier limit ($\Delta\text{E}\ll k_BT$) f$_0$ is equal to \cite{BRE-12,TAN-12,BRO-63} 
\begin{equation}
	f_0=\frac{\alpha\gamma}{1+\alpha^2}\sqrt{\frac{H_K^3M_sV}{2\pi k_BT}}\left(1-\frac{H}{H_K}\right)\left(1-\frac{H^2}{H_K^2}\right)
	\label{eq-f0}
\end{equation}
with $H_K=2K_{\text{u1}}/M_s$.\\

The probability that a particle does not switch during a certain time $\Delta T$ is given by 
$\frac{dP_\text{not}}{dt}=-fP_{\text{not}}$\cite{BRE-12}. In a constant field this leads to
\begin{equation}
	\ln{P_{\text{not}}}=\int_0^{\Delta T}f dt => P=1-\exp(-f\Delta t)
	\label{eq-prob3}
\end{equation}
The next switching time for a particle can thus be generated with 
$t=-\frac{1}{f}\ln{(1-P)}$
by generating a random number P, uniformely distributed between 0 and 1\cite{LEE-14}.
When the simulation reaches this time, the magnetization of the particle is switched to its opposite direction and a new switching time is generated.

\subsection{Time integration schemes}
\label{timestep}
Vinamax provides the user with a wide range of different methods to numerically integrate eq. (\ref{ll-eq}). \cite{BUT-08} 
\begin{itemize}
	\item Euler's method
	\item Heun's method
	\item 3th order Runge Kutta method
	\item 4th order Runge Kutta method
	\item Dormand-Prince (5th order, adaptive step)
	\item Fehlberg method (6th order, adaptive step)
	\item Fehlberg method (7th order, adaptive step)
\end{itemize}
The adaptive timestep methods incorporate a lower-order scheme, which provides an estimate of the error on the obtained results. Using eq.(\ref{eq-timestep})\cite{GUS-92}, in which $O$ stands for the order of the solver,  it is then possible to estimate an optimal timestep $dt_{\text{opt}}$ based on the error tolerance $\epsilon$ which can be set by the user. 
\begin{equation}
	dt_{\text{opt}}=dt_{\text{current}}\left(\frac{\epsilon}{dt_{\text{current}}\tau_{\text{max}}}\right)^{(1/O)}
	\label{eq-timestep}
\end{equation}

Note that adaptive timesteps can not be used when using the stochastic thermal field, as the size of this field depends on the timestep (eq. (\ref{eq-therm})), which turns eq. (\ref{eq-timestep}) unstable. Because the thermal fields should be uncorrelated in time \cite{BRO-63} the timestep should not be set smaller than 1e-13 s.

\section{Results}
\label{validation}
The use of the macrospin approximation in Vinamax has the advantage that exchange interactions do not have to be evaluated. However, this has the drawback that the micromagnetic standard problems\cite{MAG}, which do incorporate exchange interactions, can not be solved by Vinamax. In this section we therefore simulate some simple, yet general, problems and compare their solution with the micromagnetic simulation software \textsc{MuMax3}\cite{VAN-11}.\\

%The Vinamax input files, which contain a detailed description of the problems, can be found in the supplemental materials\cite{SUP}.

\subsection{Problem one: Precession and Damping}
\label{problemone}
In this first problem an isotropic single particle is considered at 0 K. It is initialised with the magnetization along the x-direction and an external field of 10 mT is applied along the z-direction. The magnetization gyrates around this axis with a frequency of 28 GHz/T and slowly damps towards the z-axis, as dictated by the Landau-Lifshitz equation.\\

\begin{figure}
%\centering
\includegraphics[width=8.6cm]{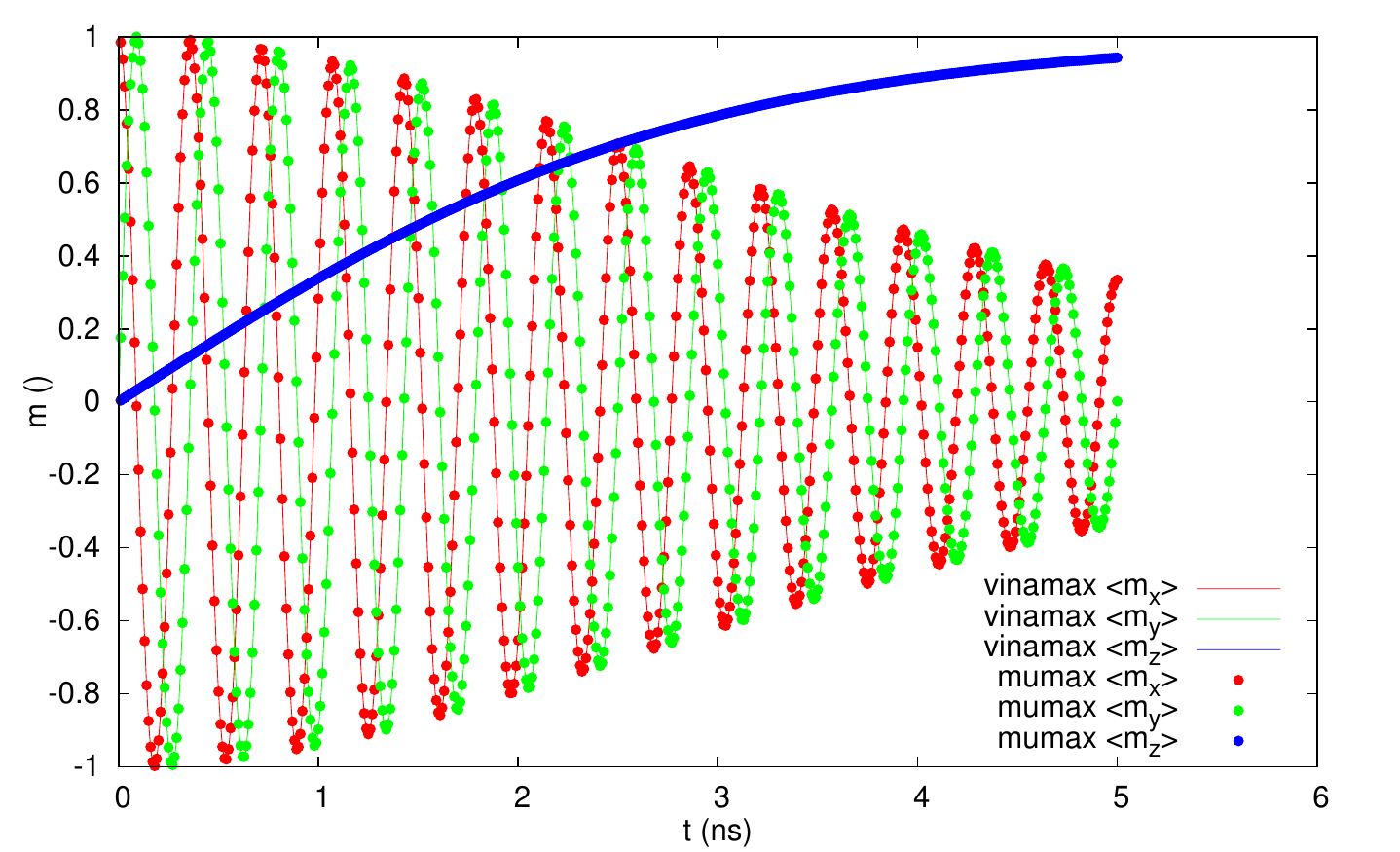}
\caption{\label{Fig_example1} The average magnetization components as a function of time for simulation problem one. The magnetization precesses around the z-axis with a frequency of 0.28 GHz and slowly damps towards this axis. The different colours denote the different magnetization components. The full lines denote the results obtained with Vinamax, while the dots correspond to the results obtained with \textsc{MuMax3}.}
\end{figure}

In Fig. \ref{Fig_example1} the results of the simulation are shown. The oscillations in the x- and y-components of the magnetization have a frequency of 0.28\,Ghz, which is in agreement with both the theoretically expected value and the result obtained with \textsc{MuMax3}. The steady increase of the z-component of the magnetization corresponds with the damping towards the z-axis ($\alpha=0.02$). 

\subsection{Problem two: Magnetostatic Interaction}
The aim of the second problem is to show that the brute force method to calculate the demagnetizing field is implemented correctly. To this end, at 0 K, two isotropic nanoparticles relax in the presence of an external field of 1 mT along the x-axis. The same simulation is also repeated without calculating the demagnetizing field to see that this problem is suited to validate the implementation; i.e. to see that the demagnetizing field has an influence.\\

The nanoparticles have a diameter of 32 nm and a saturation magnetization of 860000A/m. The damping constant $\alpha$ was set to 0.1.\\

\begin{figure}
%\centering
\includegraphics[width=8.6cm]{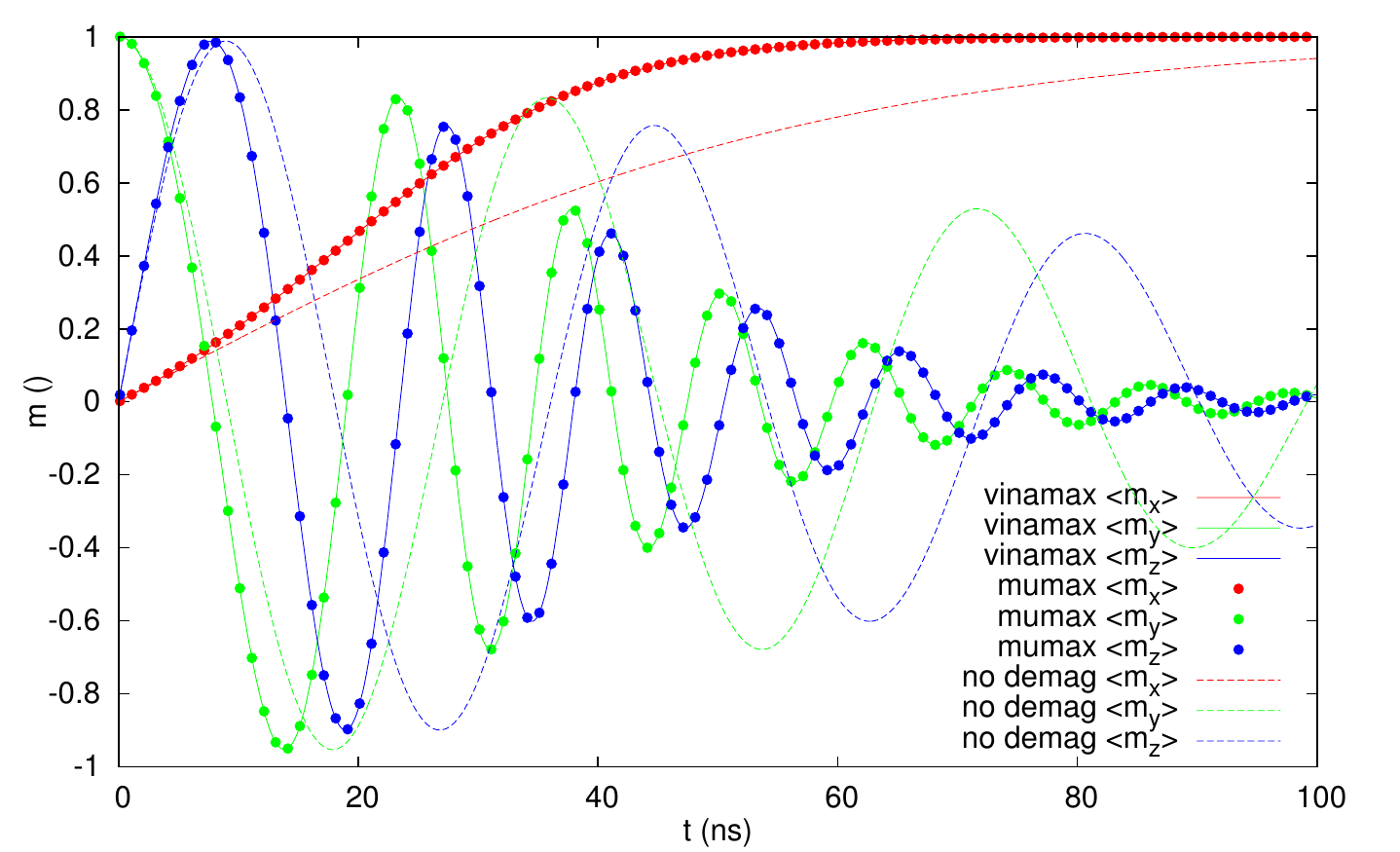}
\caption{\label{Fig_example2} The average magnetization components (different colours) as a function of time for simulation problem two. The dashed lines show the dynamics for the case in which the demagnetizing field is not included in the simulation. The full lines and big dots show the simulation results (with the demagnetizing field) obtained with Vinamax and \textsc{MuMax3} respectively. Both results are in correspondence with each other.}
\end{figure}

Again, the results are validated by comparing the results from Vinamax with those from \textsc{MuMax3}. In Fig. \ref{Fig_example2} it can be seen that the demagnetizing field is of significant importance in this problem and that the simulation results correspond with each other. 

\subsection{Problem three: Dipole Approximation method}
This example shows the agreement between the dipole approximation implementation and the brute force implementation. The same problem is also solved without taking this interaction into account so to illustrate that it is of importance in this system.\\

\begin{figure}
%\centering
\includegraphics[width=8.6cm]{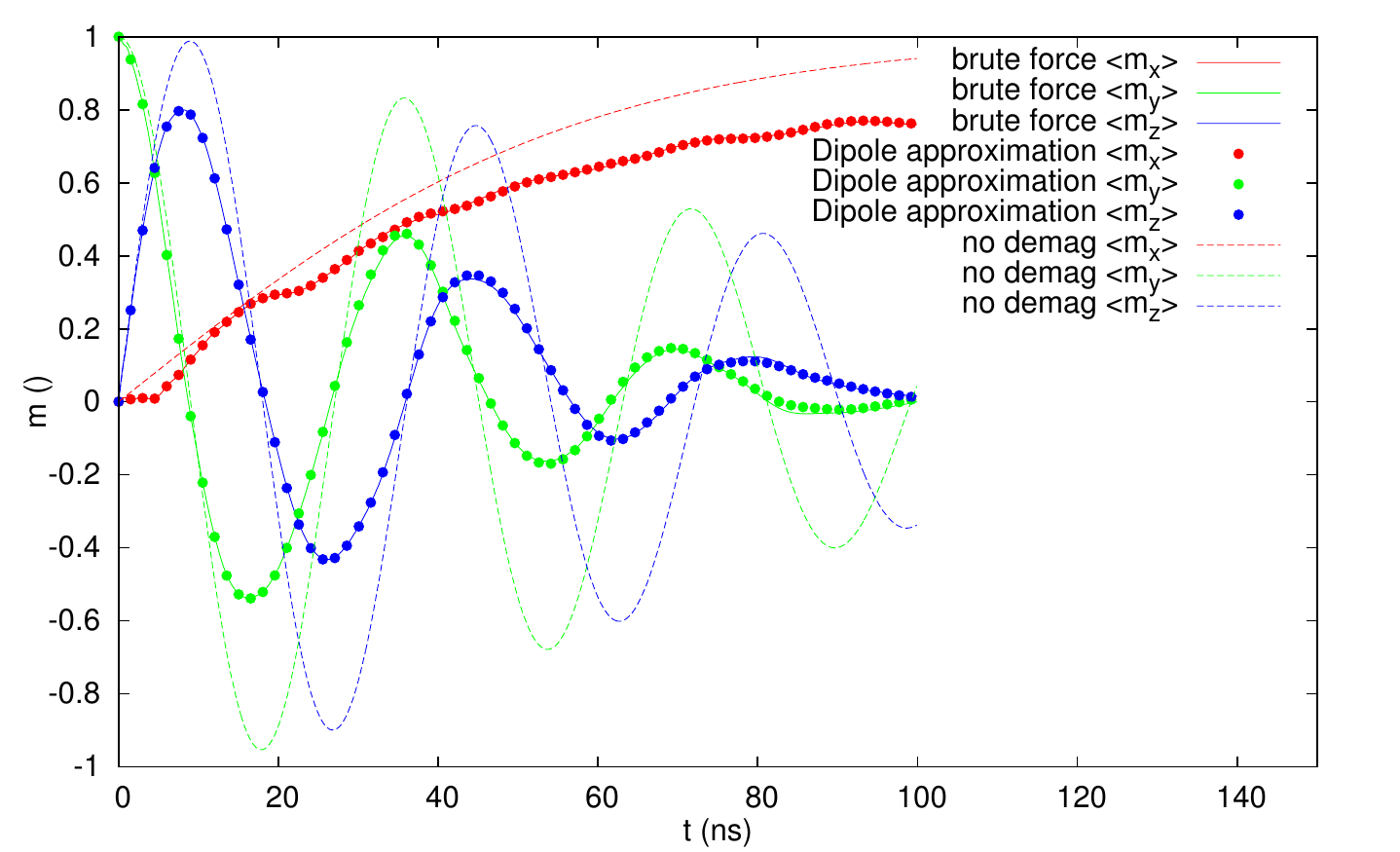}
\caption{\label{Fig_example3} The average magnetization components (different colours) as a function of time for simulation problem three. The dashed lines illustrate that the demagnetizing field has an influence in this problem. The full lines and the dots correspond with the results obtained using the brute force or dipole approximation method (with $\beta=0.4$) respectively in Vinamax, and are in correspondence with each other.}
\end{figure}

In this problem 256 isotropic nanoparticles with a diameter of 32\,nm and saturation magnetization 860000 A/m are created with a spatially uniform distribution in a cube with a side of 2\,$\mu$m with their magnetization along the z-axis. They relax at 0 K in the presence of an  external field of 1 mT along the x-axis ($\alpha=0.1$). In Fig. \ref{Fig_example3} the results of these simulations are visualised.\\

\begin{figure}
%\centering
\includegraphics[width=8.6cm]{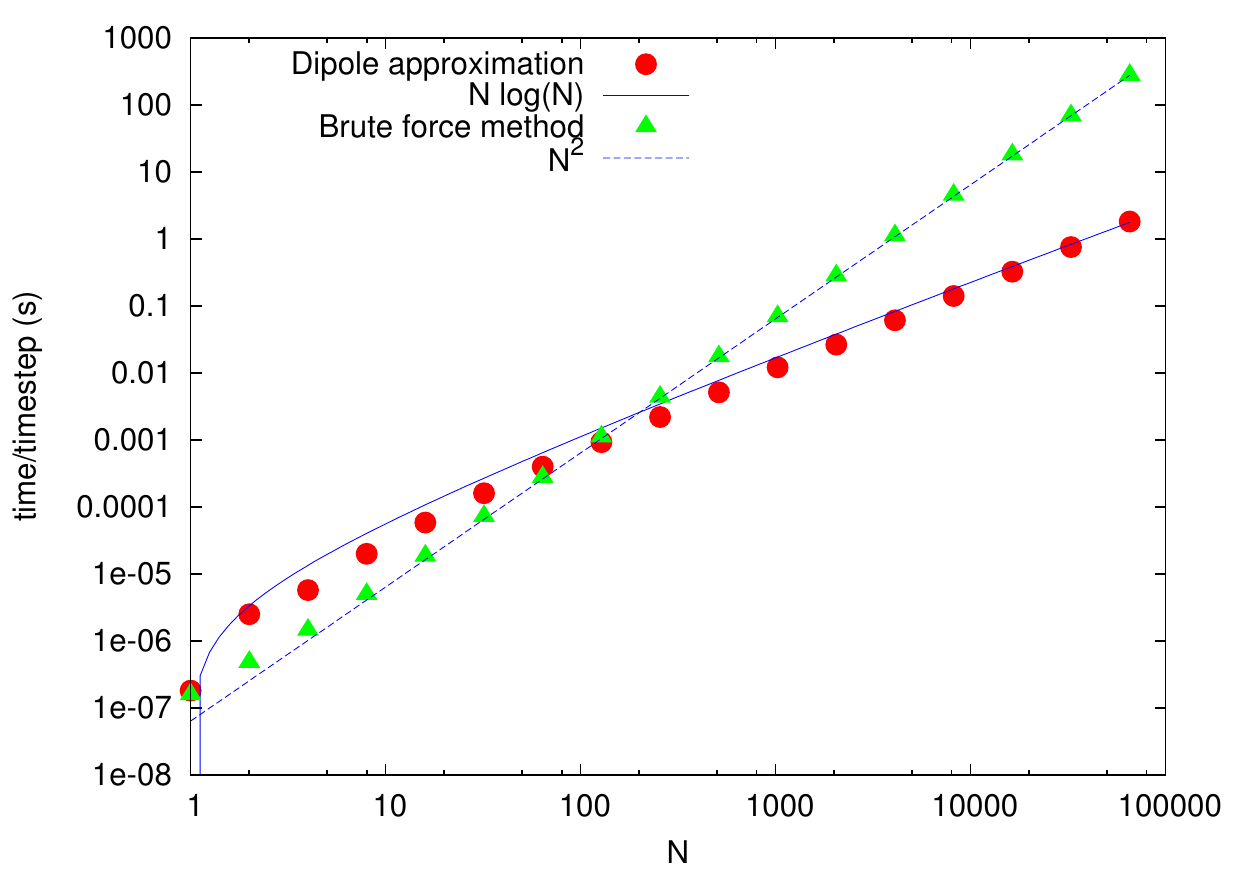}
\caption{\label{Fig_scaling} The time necessary to calculate one timestep, with Heun's solver, as function of $N$, the number of particles in the system. The green triangles correspond with the brute force evaluation of the dipole-dipole interactions, while the red dots correspond with the dipole approximation method. To visualize the scaling behaviour of both algorithms, a fit to a function $\sim N^2$(blue dots) and $\sim N\log(N)$ (blue line) is shown.}
\end{figure}
We have also investigated the performance of the different methods for evaluating the demagnetizing field in Vinamax. Fig. \ref{Fig_scaling} shows the scaling behaviour of the time it takes to calculate one timestep with the brute force evaluation of the demagnetizing field versus the dipole approximation method. All simulations were performed on an Intel Core i7-3770 CPU @ 3.40GHz. The brute force method scales as $\mathcal{O}(N^2)$. The dipole approximation method is slower for smaller numbers of particles, but scales as $\mathcal{O}(N\log(N))$. From 512 particles on, the dipole approximation method is faster than the brute force calculation and for large numbers of particles this results in an enormous  speedup of the simulation. Note that we adapted the volume of the simulation to the number of particles to keep the concentration of the particles constant.\\

\subsection{Problem four: Lognormal diameter distribution}
\label{relax}
\begin{figure}
%\centering
\includegraphics[width=8.6cm]{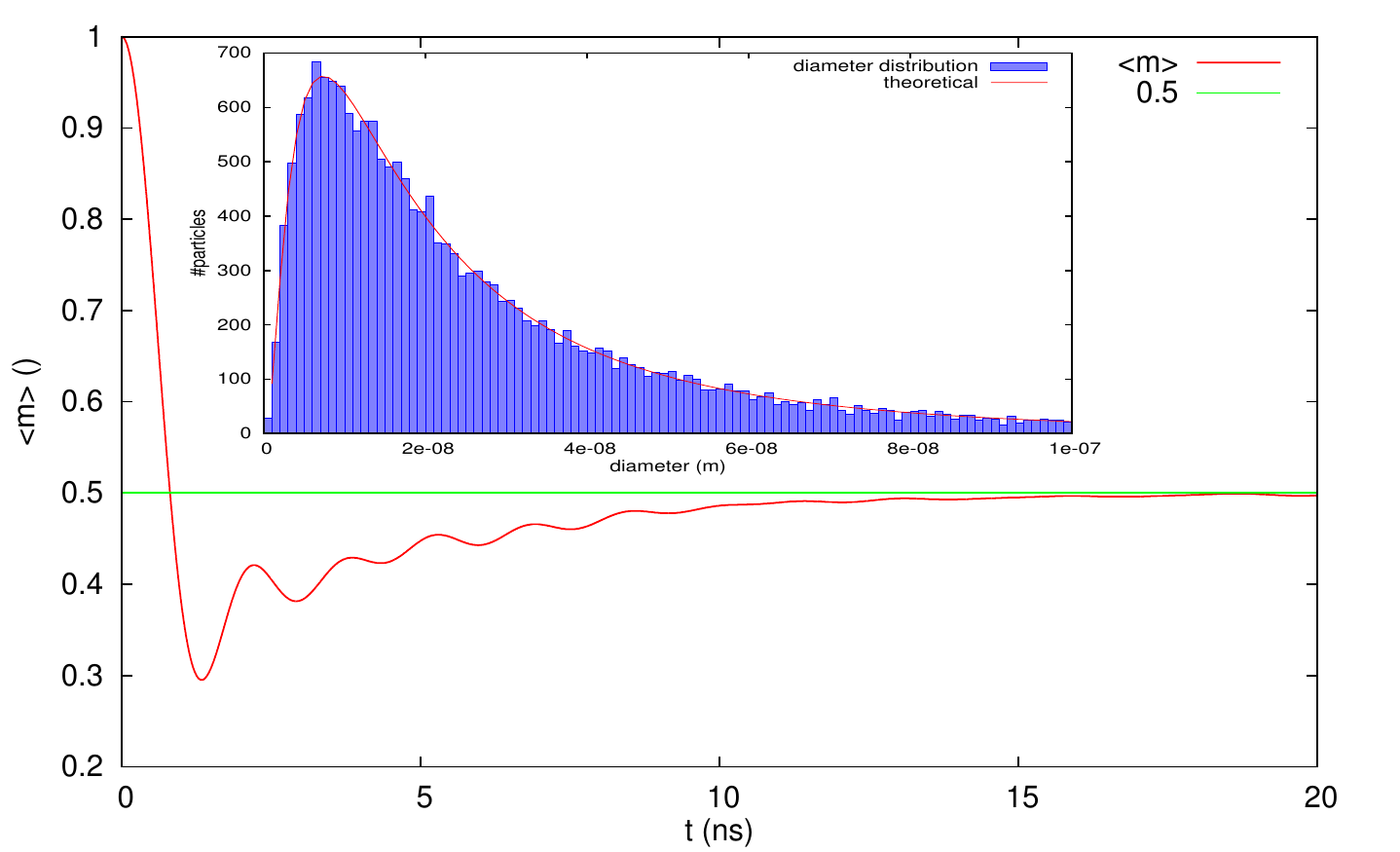}
\caption{\label{Fig_lognormal} The average magnetization $\sqrt{(m_x^2+m_y^2+m_z^2})$ versus time. The magnetization relaxes towards 0.5, as predicted analyticaly. Inset: the particle size distribution for problem four. 20000 particles were created with their diameter drawn from a lognormal distribution with $\mu=20$\,nm and $\sigma=1$\,nm. The theoretical distribution function for these values (eq. (\ref{eq-lognormal})) is shown in red to show the correspondence.}
\end{figure}

As the distribution of nanoparticle diameters $D$ often follows a lognormal distribution\cite{WIE-12,BRA-13}, the diameter of the particles in Vinamax can also be drawn from such a distribution, as shown in Fig. \ref{Fig_lognormal}. Eq. (\ref{eq-lognormal}) shows the probability density function $P(D)$ for a lognormal distribution with $\mu$ and $\sigma$ the mean and standard deviation of the logarithm of the diameter.

\begin{equation}
	P(D)=\frac{1}{\sqrt{2\pi}\sigma D}\exp{\left(-\frac{\ln^2(D/\mu)}{2\sigma^2}\right)}
	\label{eq-lognormal}
\end{equation}
Problem four shows how an ensemble of 20000 particles at 0 K, which are initially magnetized with their magnetization along the z-axis, relaxes towards a randomly chosen (per particle) anisotropy direction. The magnetization of each particle relaxes towards its anisotropy axis in the direction above the horizontal plane. After a short relaxation time, the average magnetization is thus expected to be $\int_0^{\pi/2}\sin(x)\cos(x)dx=0.5$. Fig. \ref{Fig_lognormal} shows that this is indeed the case.

%\begin{figure}
%%\centering
%\includegraphics[width=8.6cm]{example6.pdf}
%\caption{\label{Fig_example6} Average magnetization $\sqrt{(m_x^2+m_y^2+m_z^2})$ versus time}
%\end{figure}

\subsection{Application: Thermal switching time distribution and magnetization relaxation}
\label{application}
In the pevious section \ref{relax} the magnetic relaxation of an ensemble of nanoparticles was shown on the nanosecond timescale. The subject of this section is the thermal N\'eel relaxation at a much larger timescale.\\
In an ensemble of single domain particles with uniaxial anistropy, there exist two ground states along the anisotropy axis for every particle. Due to thermal fluctuations, the magnetization of each particle will switch between these two directions at random intervals. The total magnetization of the ensemble will thus slowly relax towards 0. The rate of this relaxation is given by the N\'eel relaxation time\cite{N-49} $\tau_N$, eq. (\ref{eq-neel}).  
\begin{equation}
	\tau_N=\tau_0\exp\left(\frac{K_{u1}V}{k_BT}\right)
	\label{eq-neel}
\end{equation}
In this equation,  $\tau_0$ is an extinction time with values between $10^{-8}$ and $10^{-12}$\,s \cite{WIE-12,BES-92}, and is related to $f_0$ (eq. (\ref{eq-f0})) by $\tau_0=\frac{1}{2f_0}$.

\begin{figure}
%\centering
\includegraphics[width=8.6cm]{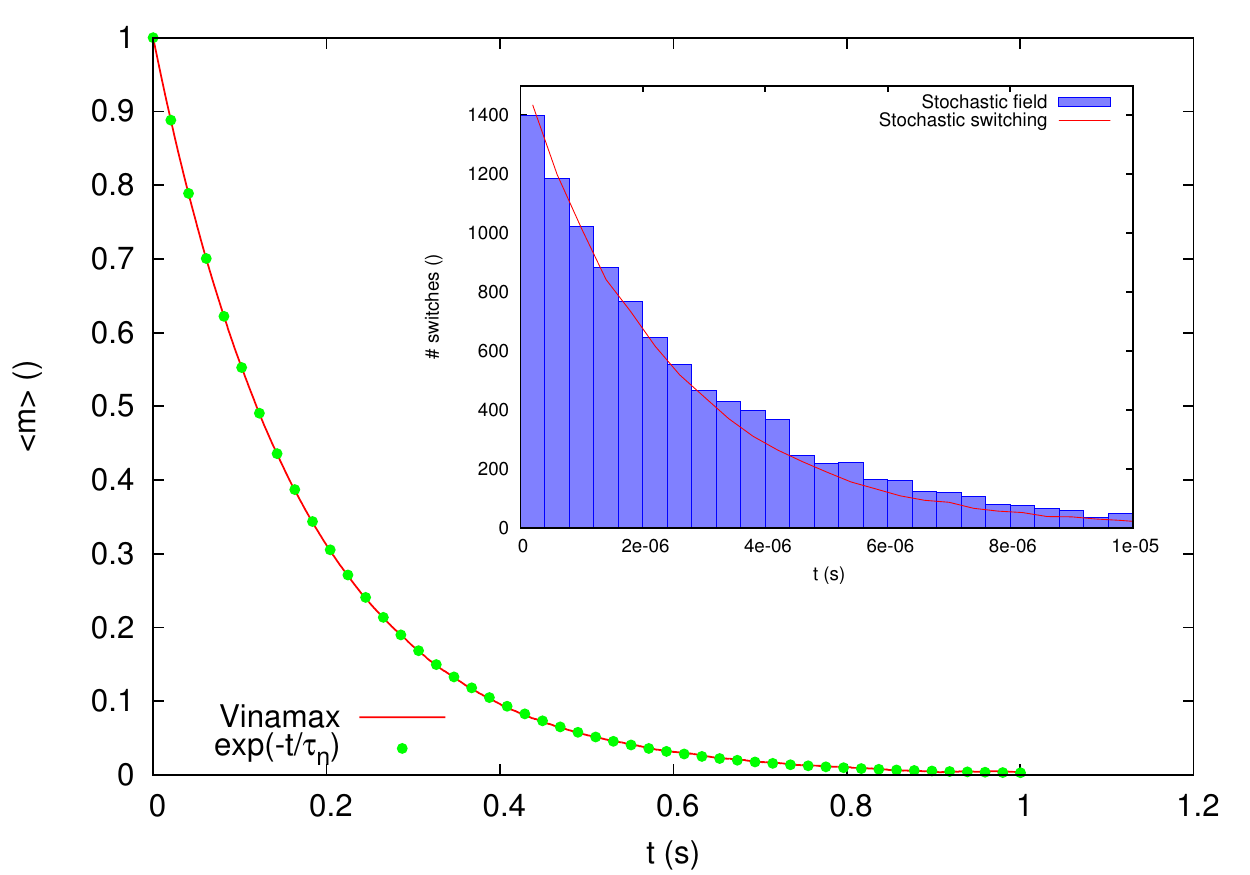}
\caption{\label{Fig_distribution} The magnetization as function of time for an ensemble of one million nanoparticles. Because of this large number, there are no fluctuations visible and the relaxation curve exactly follows the theoretical curve. The simulated nanoparticles have a saturation magnetization of 400000 A/m and uniaxial anisotropy of 10000 J/$m^3$, along the z-axis. They are simulated at a temperature of 300 K. The diameter of the particles is 25 nm and $\alpha=0.05$. The particles are placed far enough apart to neglect the demagnetizing field.
The inset shows the switching time distribution for one particle with a diameter of 18 nm for two approaches to take thermal effects into account. The particle with the stochastic field was simulated until it switched 10000 times (blue bars) and the particle with the stochastic switching (red line) was simulated until 100000 switching events were recorded (rescaled to 10000, to be able to compare both distributions). Both approaches show good agreement on the switching time distribution.}
\end{figure}

In this problem we look at the distribution of switching times of uniaxial particles with a diameter of 25 nm. As can be seen in inset Fig. \ref{Fig_distribution}, both approaches (a stochastic thermal field (section \ref{brownfield}) and stochastic switching (section \ref{jumpnoise})) show good agreement in the distribution of switching times.\\ 
The result of a lot of these switches on the magnetization is given by\cite{WIE-12}
\begin{equation}
	m=\int_{V}m_0\exp{\left(-\frac{t}{\tau_N(V)}\right)}P(V)dV
	\label{eq-relax}
\end{equation}
which equals $m= m_0\exp{\left(-\frac{t}{\tau_N}\right)}$  for particles of equal size. In Fig. \ref{Fig_distribution} the resulting magnetization of a ensemble of one million particles is shown, together with the analytical result of eq. \ref{eq-relax}.

\section{Discussion}
\label{conclusions}
The aim of this paper is to present Vinamax: a numerical software package that performs micromagnetic simulations of magnetic nanoparticles, approximated by macrospins. We have validated Vinamax in various problems against other micromagnetic software (\textsc{MuMax3}) and analytical results. Each problem (sections \ref{problemone} to \ref{relax}) shows that a different part of Vinamax is implemented correctly and the results correspond well to the expected ones. In section \ref{application} Vinamax is then used to solve a challenging problem. Due to the dipole approximation method that scales as $\mathcal{O}(N\log(N))$ with $N$ the number of particles, it is possible to simulate systems with large amounts of particles, and on large timescales. We emphasize that vinamax can be used as a research tool in biomedical applications, where we especially aim at nanoparticle imaging techniques, such as relaxometry where the collective effect of nanoparticles still is not completely understood. In this domain Vinamax could for example be used to investigate the effect of the dipolar interaction on the relaxation curves.  \\

%%%%%%%%%%%%%%%%%%%%%%%%%%%%%%%%%%%%%%%%%%%%%%%%%%%%%%%%%%%%%%%%%%%%%%%%%%%%%%%%%
%\begin{acknowledgements}
J.L. Would like to thank M. Dvornik for fruitful discussions.
This work is supported by the Flanders Research Foundation (A.V.)
%\end{acknowledgements}

\bibliographystyle{spmpsci}      % mathematics and physical sciences
%\bibliography{/home/jonathan/Documents/library/Papers/savedrecs.bib}   % name your BibTeX data base

%\begin{thebibliography}{}
% etc
%\end{thebibliography}

\end{document}